\documentclass[prl,twocolumn,nofootinbib,showpacs,amsmath,amssymb,floatfix,aps,superscriptaddress]{revtex4-1}

\usepackage{graphicx,color,framed}
\usepackage{hyperref}
\usepackage{times}
\usepackage{enumerate}
\usepackage{lipsum}
\usepackage{slashed}

\hypersetup{
    colorlinks=true, 
    linktoc=all,     
    linkcolor=blue,  
}

\def \beq {\begin{equation}}
\def \eeq {\end{equation}}
\def \beqa {\begin{eqnarray}}
\def \eeqa {\end{eqnarray}}
\def \bseq {\begin{subequations}}
\def \eseq {\end{subequations}}

\newcommand \ran {\rangle}
\newcommand \lan {\langle}

\newcommand \lam {\lambda}
\newcommand \pd {\partial}

\newcommand \mb {\mathbf}

\newcommand \nnb {\nonumber}
\newcommand \ov {\overline}

\newcommand \vphi {\varphi}

\begin{document}

\title{Semi-classical wave packet dynamics in non-uniform electric fields}

\author{Matthew F. Lapa}
\email[email address: ]{mlapa@uchicago.edu}
\affiliation{Kadanoff Center for Theoretical Physics, University of Chicago, Illinois 60637, USA}

\author{Taylor L. Hughes}
\email[email address: ]{hughest@illinois.edu}
\affiliation{Department of Physics and Institute for Condensed Matter Theory, University of Illinois at Urbana-Champaign, Urbana, IL, 61801-3080, USA}

\begin{abstract}

We study the semi-classical theory of wave packet dynamics in crystalline solids extended to include the effects of a non-uniform 
electric field. In particular, we derive a correction to the semi-classical equations of motion (EOMs) for the dynamics of the wave packet
center that depends on the gradient of the electric field and on the quantum metric 
(also called the Fubini-Study, Bures, or Bloch metric) on the
Brillouin zone. We show that the physical origin of this term is a contribution to the total energy of the
wave packet that depends on its electric quadrupole moment and on the electric field gradient. 
We also derive an equation relating the electric quadrupole moment of a sharply-peaked wave packet to the quantum 
metric evaluated at the wave packet center in reciprocal space. Finally, we explore the physical consequences of this
correction to the semi-classical EOMs. We show that in a metal with broken time-reversal and inversion
symmetry, an electric field gradient can generate a longitudinal current which 
is linear in the electric field gradient, and which depends on the quantum metric at the Fermi surface. 
We then give two examples of concrete lattice models in which this effect occurs. Our results show that non-uniform 
electric fields can be used to probe the quantum geometry of the electronic bands in metals and open the door to further 
studies of the effects of non-uniform electric fields in solids.  

\end{abstract}


\maketitle

Electron wave packets in a solid placed in an applied electric field experience an anomalous contribution to 
their velocity which has its origin in the \emph{Berry curvature} of the electronic 
bands. This anomalous velocity is responsible for the quantized Hall conductivity of Chern insulators, and the 
intrinsic contribution to the anomalous Hall effect in metals, among other things, 
and these effects can all be understood within
a framework based on the semi-classical equations of motion (EOMs) for electron wave packets in 
solids~\cite{KL,Kohn-Luttinger,luttinger1958,adams-blount,chang1995berry,chang-niu,sundaram-niu,haldane2004} (see also
the review~\cite{niu-review}). Combining the semi-classical EOMs with a Boltzmann equation approach to transport
is particularly useful in the search for novel physical consequences of band geometry and 
topology~\cite{chang1995berry,chang-niu,Moore2010,Moore2013,Moore2015,Fu2015}.

One issue which is not addressed by the current semi-classical framework is the effect of a
non-uniform electric field on wave packet motion. However, it is known that non-uniform electric fields 
probe some of the most subtle and interesting effects in condensed matter systems, for example the
Hall viscosity in quantum Hall systems~\cite{hoyos-son,bradlyn2012} and electrical multipole moments in 
insulators~\cite{benalcazar2017,wheeler2018,cho2018}. 
 In addition, it is likely that non-uniform electric fields can 
have significant effects in metals where the partially filled conduction band can respond quickly to an applied field.
These expectations motivate a systematic study of the semi-classical EOMs in an expansion in 
\emph{spatial derivatives} of the external electric field. 

In this article we initiate this study by considering the semi-classical dynamics of electron wave packets in the presence
of a constant electric field \emph{gradient}. We derive a correction to the usual 
semi-classical EOM for the time derivative of the wave packet center in real space. This correction 
depends on the \emph{gradient} of the electric field, and on the \emph{quantum metric}~\cite{provost-vallee} on the Brillouin zone (BZ)  for the electronic band
whose states are used in the construction of the wave packet. The quantum metric (also called the 
Fubini-Study metric, Bures metric, Bloch metric, etc.) has previously been studied in 
the context of band theory and the semi-classical EOMs in 
Refs.~\onlinecite{marzari-vanderbilt,matsuura2010momentum,Neupert,kolodrubetz2013classifying,legner2013relating,
ma2014quantum,gao2014field,gao2015geometrical,srivastava2015signatures,piechon2016geometric,freimuth2017geometrical,
liang2017wave,ozawa2018steady,iskin2018quantum,ozawa2018extracting,bleu2018effective,iskin2018exposing}.

The correction to the semi-classical EOMs that we derive depends on the \emph{derivative}
of the quantum metric. As a consequence, we show that this correction does not affect transport in insulators. 
On the other hand, we show one direct effect of this correction on transport in metals where it can lead
to a longitudinal current proportional to the electric field gradient and to the quantum metric at
the Fermi surface. Thus, our result shows that the quantum geometry of bands in metals can be probed by a 
transport experiment using a non-uniform electric field. We now turn to an explanation of our results.

\textit{Setup:} We study the dynamics of electrons of charge $Q=-e$ in a crystal and in the presence of a 
\emph{time-independent} electric field $\mb{E}(\mb{x})$. 
Let $\hat{x}^{\mu}$, $\hat{p}_{\nu}$, $\mu,\nu=1,\dots,D$, denote the position
and momentum operators for a single electron in $D$ spatial dimensions, with 
$[\hat{x}^{\mu},\hat{p}_{\nu}]=i\hbar\delta^{\mu}_{\nu}$. The single particle Hamiltonian is
\beq
	\hat{H}=\hat{H}_0 + Q\vphi(\hat{\mb{x}})\ ,
\eeq
where $\hat{H}_0$ is a Hamiltonian for an electron in a periodic potential $V(\hat{\mb{x}})$, for example the 
standard non-relativistic Hamiltonian 
$\hat{H}_0= \frac{1}{2m}\delta^{\mu\nu}\hat{p}_{\mu}\hat{p}_{\nu} + V(\hat{\mb{x}})$
for particles of mass $m$. In fact, our only requirement for $\hat{H}_0$ is that it be subject to Bloch's theorem. 
The second term in $\hat{H}$ captures the coupling to the electric field 
$\mb{E}(\mb{x})$, which is determined by the potential $\vphi(\mb{x})$ as 
$E_{\mu}(\mb{x})= -\frac{\pd \vphi(\mb{x})}{\pd x^{\mu}}$. 

Bloch's theorem implies that $\hat{H}_0$ has a basis of eigenstates (``Bloch waves") $|\psi_{n,\mb{q}}\ran$ which are labeled 
by a band index $n$ and a wavevector $\mb{q}$ (in the first BZ) and obey 
$\hat{H}_0|\psi_{n,\mb{q}}\ran= \mathcal{E}_n(\mb{q})|\psi_{n,\mb{q}}\ran$, where 
$\mathcal{E}_n(\mb{q})$ are the energy eigenvalues. In addition, we can write
$|\psi_{n,\mb{q}}\ran= e^{i q_{\mu}\hat{x}^{\mu}}|u_{n,\mb{q}}\ran$ where the function
$u_{n,\mb{q}}(\mb{x}):= \lan\mb{x}|u_{n,\mb{q}}\ran$ has the periodicity of the crystal lattice. 
Note that the Bloch states are time-independent since we have made the
simplifying assumption that our Hamiltonian is time-independent.
We normalize the Bloch states so that $\lan \psi_{n,\mb{q}}|\psi_{n',\mb{q}'}\ran= \delta_{n,n'}\delta^{(D)}(\mb{q}-\mb{q}')$,
which implies that $\lan u_{n,\mb{q}}| u_{n',\mb{q}}\ran = \delta_{n,n'}$. Here, the
the inner product of the $|u_{n,\mb{q}}\ran$ is defined as integration over the real space unit cell times a factor of 
$\frac{(2\pi)^D}{v_c}$, where $v_c$ is the volume of the real space unit cell~\cite{blount}.
We also introduce a crystal momentum operator $\hat{q}_{\mu}$ which is diagonal
in the basis of Bloch states and satisfies $\hat{q}_{\mu}|\psi_{n,\mb{q}}\ran= q_{\mu}|\psi_{n,\mb{q}}\ran$.

We are interested in the leading corrections to the semi-classical EOMs due to a non-zero electric field gradient,
and so we choose a potential of the form 
$\vphi(\mb{x})= -E^{(0)}_{\mu}x^{\mu}-\frac{1}{2}E^{(0)}_{\mu\nu}x^{\mu}x^{\nu}$,
where $E^{(0)}_{\mu}$ and $E^{(0)}_{\mu\nu}$ (with $E^{(0)}_{\mu\nu}=E^{(0)}_{\nu\mu}$) are 
two sets of constant parameters. The components of the electric field are then 
$E_{\mu}(\mb{x})= E^{(0)}_{\mu} + E^{(0)}_{\mu\nu}x^{\nu}$.
We see that $E^{(0)}_{\mu}$ specify the uniform part of the electric field, while $E^{(0)}_{\mu\nu}$ specify the 
electric field \emph{gradient}.

\textit{Wave packets and their first moments:} We study the time-evolution (using the full Hamiltonian $\hat{H}$) 
of a wave packet $|\Psi(t)\ran$ constructed from the Bloch states 
$|\psi_{n,\mb{q}}\ran$. We assume that the wave packet is constructed from states within a single band,
and so we drop the band index $n$ from the notation. We define this wave packet state as 
$|\Psi(t)\ran= \int d^D\mb{q}\ a(\mb{q},t)|\psi_{\mb{q}}\ran$,
where $a(\mb{q},t)$ is a complex amplitude which must satisfy the normalization condition 
$\int d^D\mb{q}\ |a(\mb{q},t)|^2=1$ ($\mb{q}$ integrals run
over the first BZ). By plugging into the Schrodinger equation 
$i\hbar\frac{d}{dt}|\Psi(t)\ran = \hat{H}|\Psi(t)\ran$, one can show that $a(\mb{q},t)$ satisfies 
\begin{align}
	i\hbar\dot{a}(\mb{q},t)= a(\mb{q},t)\mathcal{E}(\mb{q}) + Q \int d^D\mb{q}'\ a(\mb{q}',t) \lan\psi_{\mb{q}}|\vphi(\hat{\mb{x}})|\psi_{\mb{q}'}\ran\ , \label{eq:a-eqn}
\end{align}
where the dot denotes a time derivative. 

The semi-classical EOMs for wave packet dynamics in solids can be derived by studying the dynamics
of the first \emph{moments} $X^{\mu}(t)$ and $K_{\mu}(t)$ of the wave packet in position and reciprocal space, 
respectively. These are defined by $X^{\mu}(t) = \lan\Psi(t)|\hat{x}^{\mu}|\Psi(t)\ran$ and
$K_{\mu}(t) = \lan\Psi(t)|\hat{q}_{\mu}|\Psi(t)\ran= \int d^D\mb{q}\ q_{\mu} |a(\mb{q},t)|^2$.
We derive the semi-classical EOMs for $X^{\mu}(t)$ and $K_{\mu}(t)$ by first computing the 
\emph{exact} expressions for $\dot{X}^{\mu}(t)$ and $\dot{K}_{\mu}(t)$, and then truncating these expressions using the
assumption that the wave packet is sharply-peaked about the locations $X^{\mu}(t)$ and $K_{\mu}(t)$ in position and 
reciprocal space. To derive the equations for $\dot{X}^{\mu}(t)$ and $\dot{K}_{\mu}(t)$ we simply 
differentiate the expressions for $X^{\mu}(t)$ and $K_{\mu}(t)$ with respect to time, and then we substitute in 
for $\dot{a}(\mb{q},t)$ and $\dot{\ov{a}}(\mb{q},t)$ using Eq.~\eqref{eq:a-eqn} and its complex conjugate.

After a tedious but straightforward calculation, we find that the equation for 
$\dot{X}^{\mu}(t)$ takes the form
\begin{widetext}
\begin{align}
	\dot{X}^{\mu}(t) =\frac{1}{\hbar} \Big\lan\frac{\pd \mathcal{E}(\hat{\mb{q}})}{\pd q_{\mu}}\Big\ran_t - \frac{1}{\hbar}QE^{(0)}_{\nu} \Big\lan\Omega^{\mu\nu}(\hat{\mb{q}})\Big\ran_t  -\frac{1}{2\hbar}Q E^{(0)}_{\nu\lam}\Big\lan\{ \hat{x}^{\lam}, \Omega^{\mu\nu}(\hat{\mb{q}})\}\Big\ran_t - \frac{1}{2\hbar}Q E^{(0)}_{\nu\lam} \Big\lan \frac{\pd g^{\nu\lam}(\hat{\mb{q}})}{\pd q_{\mu}}\Big\ran_t\ , \label{eq:X-dot-eqn}
\end{align}
\end{widetext}
where $\lan \cdot \ran_t$ denotes an expectation value in the state $|\Psi(t)\ran$, and
$\{\cdot,\cdot\}$ denotes an anti-commutator (third term on the right-hand side). In this equation $\Omega^{\mu\nu}(\mb{q})$
is the Berry curvature, which is expressed in terms of the Berry connection 
$\mathcal{A}^{\mu}(\mb{q})= i\Big\lan u_{\mb{q}}\Big| \frac{\pd u_{\mb{q}}}{\pd q_{\mu}}\Big\ran$ as 
$\Omega^{\mu\nu}(\mb{q}) = \frac{\pd\mathcal{A}^{\nu}(\mb{q})}{\pd q_{\mu}}-\frac{\pd\mathcal{A}^{\mu}(\mb{q})}{\pd q_{\nu}}$. The quantity $g^{\mu\nu}(\mb{q})$ is the \emph{quantum metric} on the BZ, and is defined as
\begin{align}
	g^{\mu\nu}(\mb{q})=\frac{1}{2}\Bigg( \Big\lan \frac{\pd u_{\mb{q}}}{\pd q_{\mu}}\Big| \frac{\pd u_{\mb{q}}}{\pd q_{\nu}}\Big\ran \nnb &- \Big\lan \frac{\pd u_{\mb{q}}}{\pd q_{\mu}}\Big|u_{\mb{q}}\Big\ran \Big\lan u_{\mb{q}}\Big| \frac{\pd u_{\mb{q}}}{\pd q_{\nu}}\Big\ran \nnb \\
&+ (\mu\leftrightarrow\nu) \Bigg)\ .
\end{align}
Both $\Omega^{\mu\nu}(\mb{q})$ and $g^{\mu\nu}(\mb{q})$ are invariant under a gauge transformation 
$|u_{\mb{q}}\ran\to e^{-i f(\mb{q})}|u_{\mb{q}}\ran$ for any function $f(\mb{q})$. The equation for 
$\dot{K}_{\mu}(t)$ is much simpler, and it takes the form
\beq
	\dot{K}_{\mu}(t)= \frac{1}{\hbar}Q E_{\mu}(\mb{X}(t))\ , \label{eq:K-dot-eqn}
\eeq
where $E_{\mu}(\mb{X}(t))= E^{(0)}_{\mu} + E^{(0)}_{\mu\nu}X^{\nu}(t)$ is the electric field at the location of the
first moment $X^{\mu}(t)$.

To derive these equations, it is necessary to use explicit expressions for the matrix elements 
$\lan\psi_{\mb{q}}|\hat{x}^{\mu}|\psi_{\mb{q}'}\ran$ and $\lan\psi_{\mb{q}}|\hat{x}^{\mu}\hat{x}^{\nu}|\psi_{\mb{q}'}\ran$
of the position operator in the Bloch states. We record these expressions in Eqs.~(3) and (4) of the Supplemental Material~\cite{supp}.
In the derivation we also used several integrations by parts in integrals over the BZ, and we neglected boundary 
terms. If the amplitudes $a(\mb{q},t)$ or the Berry connection $\mathcal{A}^{\mu}(\mb{q})$ are not single-valued, then there could be 
some interesting, subtle additions to these modified EOMs. 
In the Supplemental Material we show that by a suitable choice of gauge for the
Bloch states $|\psi_{\mb{q}}\ran$, we can make $a(\mb{q},t)$ single-valued for all $t$. In that case the only possible source of
boundary corrections is the Berry connection. Here we assume that no boundary corrections arise, and we
leave a detailed discussion of any alternatives to future work.

To obtain the semi-classical EOMs for 
$X^{\mu}(t)$ and $K_{\mu}(t)$ we make the substitutions
$\hat{x}^{\mu}\to X^{\mu}(t)$ and $\hat{q}_{\mu} \to K_{\mu}(t)$ in all
expectation values in Eq.~\eqref{eq:X-dot-eqn} and Eq.~\eqref{eq:K-dot-eqn}. Our result, which is one of the main results of 
this article, is that the semi-classical EOMs take the form
\begin{subequations}
\label{eq:new-semiclassics}
\begin{align}
	\dot{X}^{\mu} &= \frac{1}{\hbar}\frac{\pd \mathcal{E}(\mb{K})}{\pd K_{\mu}} -\Omega^{\mu\nu}(\mb{K})\dot{K}_{\nu}-\frac{1}{2\hbar}QE^{(0)}_{\nu\lam}\frac{\pd g^{\nu\lam}(\mb{K})}{\pd K_{\mu}} \\
	\dot{K}_{\mu} &= \frac{1}{\hbar}Q E_{\mu}(\mb{X})\ ,
\end{align}
\end{subequations}
where we also used the second equation to rewrite part of the $\dot{X}^{\mu}(t)$ equation in terms of 
$\dot{K}_{\mu}(t)$. The main difference compared to the usual semi-classical EOMs is the term
$-\frac{1}{2\hbar}Q E^{(0)}_{\nu\lam}\frac{\pd g^{\nu\lam}(\mb{K})}{\pd K_{\mu}}$.
This new term depends on the \emph{gradient} of the electric field, since it depends on $E^{(0)}_{\nu\lam}$ but not
$E^{(0)}_{\mu}$, and it also probes the \emph{geometry} of the band 
structure since it involves the quantum metric $g^{\nu\lam}(\mb{K})$. 

\textit{Interpretation:} We now show that the new term in 
\eqref{eq:new-semiclassics} arises from an electric field-induced correction to the energy of the wave packet. 
In the absence of an electric field we have
$\lan\Psi(t)|\hat{H}_0|\Psi(t)\ran = \int d^D\mb{q}\ |a(\mb{q},t)|^2\mathcal{E}(\mb{q}) \approx \mathcal{E}(\mb{K})$, 
where $\mb{K}(t)$ is the wave packet center. In the presence of the electric field, we show in the 
Supplemental Material that the wave packet energy takes the form
\begin{align}
	\lan\Psi(t)|\hat{H}|\Psi(t)\ran 	&\approx \mathcal{E}(\mb{K}) - Q E^{(0)}_{\mu}X^{\mu} \nnb \\
	&- \frac{1}{2}Q E^{(0)}_{\mu\nu}\Bigg( X^{\mu}X^{\nu} + g^{\mu\nu}(\mb{K}) \Bigg)\ \\
	&\equiv \mathcal{E}_{eff}(\mb{X},\mb{K})\ .
\end{align}
As a result, the corrected semi-classical EOM for $\dot{X}^{\mu}(t)$ can be rewritten as
\beq
	\dot{X}^{\mu} = \frac{1}{\hbar}\frac{\pd \mathcal{E}_{eff}(\mb{X},\mb{K})}{\pd K_{\mu}} -\Omega^{\mu\nu}(\mb{K})\dot{K}_{\nu}\ .
\eeq 
We can also rewrite the equation for $\dot{K}_{\mu}(t)$ as 
$\dot{K}_{\mu} = -\frac{1}{\hbar}\frac{\pd \mathcal{E}_{eff}(\mb{X},\mb{K})}{\pd X^{\mu}}$.

In this form, the correction to the $\dot{X}^{\mu}(t)$ and $\dot{K}_{\mu}(t)$ equations closely 
resembles a similar correction which occurs for electrons 
in a magnetic field. In that case the correction to the wave packet energy arises from the magnetic moment of the 
wave packet~\cite{chang-niu}. In the present case of a non-uniform electric field, the corrections to the energy depend on 
the dipole moment $X^{\mu}(t)$ of the wave packet (the term proportional to $E^{(0)}_{\mu}$), and on the
\emph{quadrupole} moment of the wave packet (the term proportional to $E^{(0)}_{\mu\nu}$). Indeed, in the 
Supplemental Material we show that for a wave packet $|\Psi(t)\ran$ sharply peaked at position $\mb{K}$ in reciprocal space, the 
quadrupole moment is given by
\beq
	\lan \Psi(t)|\hat{x}^{\mu}\hat{x}^{\nu}|\Psi(t)\ran\approx X^{\mu}X^{\nu} + g^{\mu\nu}(\mb{K})\ . \label{eq:quadrupole}
\eeq
The correction due to the dipole moment is already present in the case of a uniform electric field, and it does not alter the
semi-classical EOMs. On the other hand, the correction proportional to the quadrupole moment is only present in a non-uniform 
field, and it does alter the semi-classical EOMs. We also note that to find
the $g^{\mu\nu}(\mb{K})$ term in $\mathcal{E}_{eff}(\mb{X},\mb{K})$, we need to expand
$Q\lan\Psi(t)|\vphi(\hat{\mb{x}})|\Psi(t)\ran$ to \emph{second order} about the wave packet center in real space, and so this
term cannot be found using the first order expansion of Ref.~\onlinecite{sundaram-niu}.

\textit{Physical consequences:} We now discuss physical consequences of the new term in Eq.~\eqref{eq:new-semiclassics} for
transport in solids. Within the semi-classical approach, the current density $j^{\mu}(\mb{r})$ at position $\mb{r}$ in the
material is given by $j^{\mu}(\mb{r})= Q\int d^D\mb{X} \frac{d^D\mb{K}}{(2\pi)^D}\ f(\mb{X},\mb{K},t)\dot{X}^{\mu} \delta^{(D)}(\mb{X}-\mb{r})$, where $f(\mb{X},\mb{K},t)$ is the non-equilibrium distribution function which specifies the occupation, at
time $t$, of the volume element $d^D\mb{X} \frac{d^D\mb{K}}{(2\pi)^D}$ at position $(\mb{X},\mb{K})$ in phase space.
The full distribution function can be obtained by solving the Boltzmann equation. In the relaxation time approximation, 
with relaxation time $\tau$, $f(\mb{X},\mb{K},t)$ takes the form of a power series in $\tau$, 
$f(\mb{X},\mb{K},t)= f_0(\mb{K}) + O(\tau)$, where $f_0(\mb{K})$ is the equilibrium distribution function specifying
the occupied states in reciprocal space at temperature $T$~\cite{chang1995berry,chang-niu,Moore2010,Moore2013,Moore2015,Fu2015}. 
In what follows, we will be interested in the currents which come from 
this zeroth order contribution, which captures the intrinsic part of the linear response of the system to the applied electric field. 
The zeroth order contribution to the current is then $j^{\mu}_0(\mb{r})= Q\int \frac{d^D\mb{K}}{(2\pi)^D}\ f_0(\mb{K})
\dot{X}^{\mu}\Big|_{\mb{X}=\mb{r}}$. In $D=2$, for example, $j^{\mu}_0(\mb{r})$ contains the intrinsic contribution 
to the anomalous Hall effect. Using Eq.~\eqref{eq:new-semiclassics}, we find that 
$j^{\mu}_0(\mb{r})$ contains the additional term
\beq
j^{\mu}_{\text{geom.}}(\mb{r})= -\frac{Q^2}{2\hbar} \int \frac{d^D\mb{K}}{(2\pi)^D}\ f_0(\mb{K}) E^{(0)}_{\nu\lam}\frac{\pd g^{\nu\lam}(\mb{K})}{\pd K_{\mu}}\ , \label{eq:j-geom}
\eeq
which involves the electric field gradient and the quantum metric. We will refer to $j^{\mu}_{\text{geom.}}$ as the 
\emph{geometric} current.

The geometric current is easiest to understand in the case of a metal in $D=1$ dimension 
(so $\mu,\nu=1$ in all equations). Recall that we considered wave packets constructed from states in a 
single band. We assume a partial filling of this band such that the Fermi level $\mathcal{E}_F$ crosses the band at the set of wave 
numbers $\{k_{I,+},k_{I,-}\}_{I\in\{1,\dots,n_F\}}$ for some integer $n_F$ (so $2n_F$ is the total number of Fermi points). Our 
notation means that $\frac{\pd \mathcal{E}(K_1)}{\pd K_1}$ is positive at a $+$ Fermi point and negative at a $-$ Fermi point 
(we assume that $\mathcal{E}_F$ is chosen so that there is no Fermi point where $\frac{\pd \mathcal{E}(K_1)}{\pd K_1}$ 
vanishes). At temperature $T=0$ the distribution function $f_0(K_1)$ is equal to $1$ if $\mathcal{E}(K_1)\leq \mathcal{E}_F$
and zero otherwise. After an integration by parts, and using
$\frac{\pd f_0(K_1)}{\pd K_1}= \sum_{I=1}^{n_F}\left[ \delta(K_1-K_{I,-}) - \delta(K_1-K_{I,+})\right]$, 
we find that ($h=2\pi\hbar$)
\beq
	j^1_{\text{geom.}} = -\frac{1}{2}\frac{Q^2}{h}E^{(0)}_{11}\sum_{I=1}^{n_F}\left[ g^{11}(K_{I,+})- g^{11}(K_{I,-})\right]\ , \label{eq:1D-current}
\eeq
which is non-zero if the sum does not equal zero. 

\begin{figure}[t]
  \centering
    \includegraphics[width= .32\textwidth]{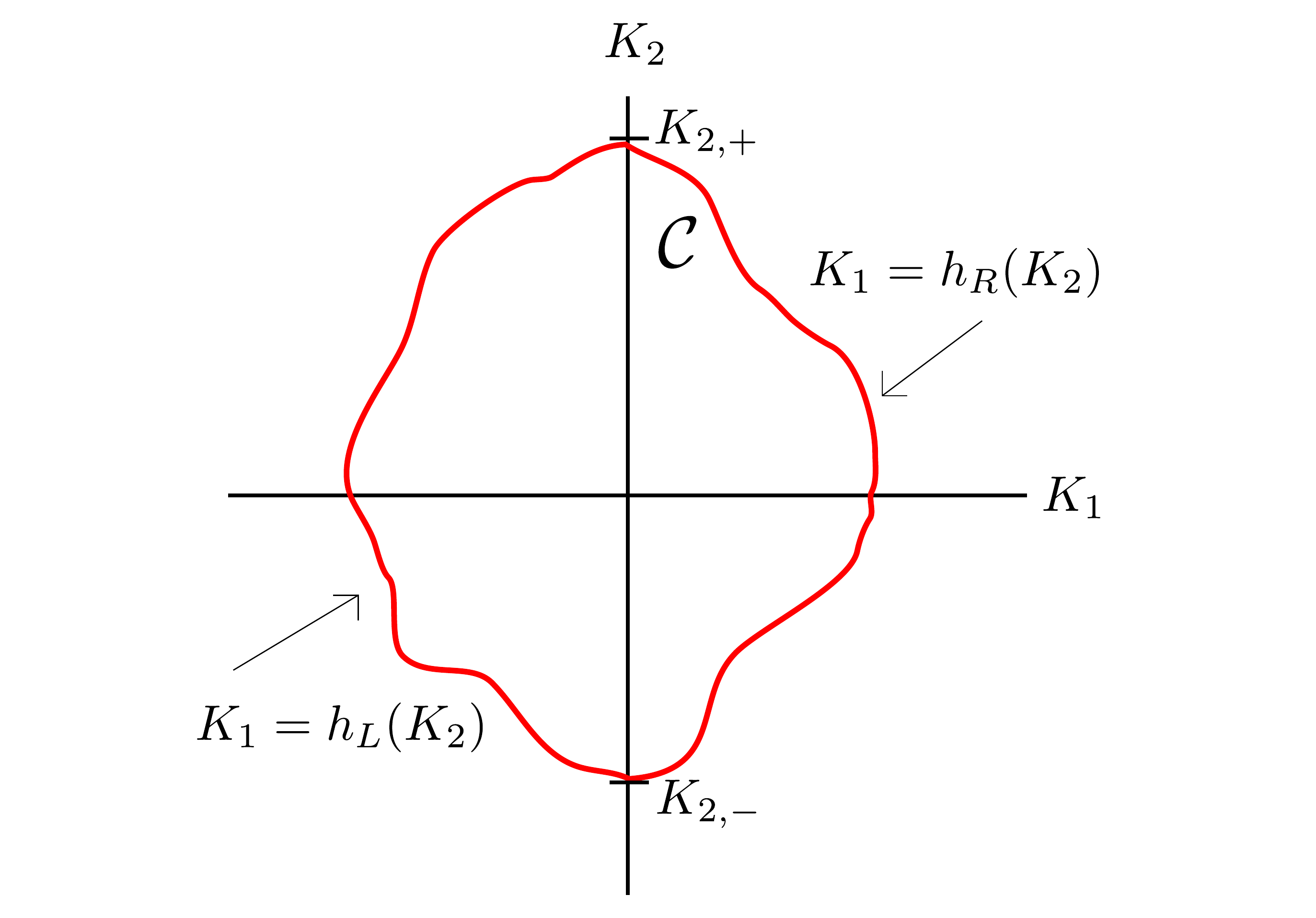} 
\caption{A Fermi surface $\mathcal{C}$ (red contour), whose segments to the left and right of the $K_2$ axis are
defined by the relations $K_1=h_L(K_2)$, $K_1=h_R(K_2)$, respectively, where $h_{R/L}(K_2)$ are single-valued functions
of $K_2$.}
\label{fig2}
\end{figure}

Next, we consider a similar example for a metal in $D=2$. To illustrate the nontrivial response
we compute $j^1_{\text{geom.}}$ as an example. We again consider
a single band and we assume the Fermi surface consists of a single 
closed contour $\mathcal{C}$. For simplicity, we assume further that the parts of $\mathcal{C}$ to the left and right of the
$K_2$ axis can be specified by single-valued functions $h_L(K_2)$, $h_R(K_2)$, such that $K_1=h_L(K_2)$ defines
the part of $\mathcal{C}$ to the left of the $K_2$ axis, and $K_1=h_R(K_2)$ defines the part to the right (note that
for a generic $\mathcal{C}$ the functions $h_{R/L}(K_2)$ would not be single-valued). Let
$K_{2,+} >0$ and $K_{2,-}<0$ be the two points where $\mathcal{C}$ intersects the $K_2$ axis. This situation
is illustrated in Fig.~\ref{fig2}. At $T=0$ the distribution function for this metal is $f_0(\mb{K})=1$ for
$\mb{K}$ inside $\mathcal{C}$, and zero otherwise.
We then have $\frac{\pd f_0(\mb{K})}{\pd K_1}= \delta(K_1-h_L(K_2))-\delta(K_1-h_R(K_2))$, and so 
\begin{align}
	j^1_{\text{geom.}}&= -\frac{Q^2}{2 h}\frac{E^{(0)}_{\nu\lam}}{2\pi}\Bigg( \int_{K_{2,-}}^{K_{2,+}} dK_2\  g^{\nu\lam}(\mb{K})\big|_{K_1=h_R(K_2)}  \nnb \\
	&- \int_{K_{2,-}}^{K_{2,+}} dK_2\ g^{\nu\lam}(\mb{K})\big|_{K_1=h_L(K_2)}\Bigg)\ . \label{eq:2D-example}
\end{align}
Since $j^1_{\text{geom.}}$ involves an integral of $g^{\mu\nu}(\mb{K})$ only on $\mathcal{C}$, we see that this current is a Fermi 
surface property, like the intrinsic contribution to the anomalous Hall effect~\cite{haldane2004}, and it vanishes in insulators (which have
a full band, $f_0(\mb{K})=1$ $\forall\ \mb{K}$).

\textit{Symmetry analysis:} In systems with time-reversal symmetry we have
$\mathcal{E}(\mb{K})=\mathcal{E}(-\mb{K})$ and $g^{\mu\nu}(\mb{K})= g^{\mu\nu}(-\mb{K})$, and identical
conditions hold in the case of inversion symmetry. These conditions imply that $j^{\mu}_{\text{geom.}}=0$.
To prove this we use these conditions to first replace $g^{\mu\nu}(\mb{K})$ in Eq.~\eqref{eq:j-geom} with
$g^{\mu\nu}(-\mb{K})$. Next, we use the fact that $f_0(\mb{K})=f_0(-\mb{K})$ if $f_0(\mb{K})$ is a function of 
$\mathcal{E}(\mb{K})$ only (as it would be in thermal equilibrium or at $T=0$). Finally, we 
change integration variables from $\mb{K}$ to $-\mb{K}$ to find that time-reversal
or inversion symmetry imply that $j^{\mu}_{\text{geom.}}= - j^{\mu}_{\text{geom.}}$, and so 
$j^{\mu}_{\text{geom.}}=0$. Therefore we must break these symmetries to obtain $j^{\mu}_{\text{geom.}}\neq 0$.

\textit{Examples in $D=1$ and $D=2$:} We now discuss two examples of lattice models of metals in $D=1$ and $D=2$
which yield a non-zero geometric current. We present the detailed results for the geometric current in these models in 
the Supplemental Material. In $D=1$ we consider the two-band model with Bloch Hamiltonian 
\beq
	H_{1D}(k)= A\sin(k)\mathbb{I} + \sin(k)\sigma^x + (m+1-\cos(k))\sigma^z\ , \label{eq:example-ham}
\eeq
where $\mathbb{I}$ is the $2\times 2$ identity matrix and $\sigma^{x,y,z}$ are the Pauli matrices. In $D=2$ we consider
the two-band model with Bloch Hamiltonian 
\begin{align}
	H_{2D}(\mb{k})&= A\sin(k_1)\mathbb{I} + \sin(k_1)\sigma^x + \sin(k_2)\sigma^y \nnb \\
	&+ (m+2-\cos(k_1)-\cos(k_2))\sigma^z \ . \label{eq:example-ham-2D}
\end{align}
In both cases we choose the parameters $m$ and $A$ so that there is an energy gap between the two bands of the model. We
then fill the lower band and partially fill the upper band to a Fermi energy
$\mathcal{E}_F$ to obtain a model of a metal. In the Supplemental Material we show that under these conditions, and when the parameter
$A\neq 0$, both of these models display a nontrivial geometric current in the presence of a non-uniform electric field in the $X^1$ 
direction (i.e., $E^{(0)}_{11}\neq 0$). In both cases the condition $A\neq 0$ is required to break inversion and/or time-reversal 
symmetry, which then allows for a non-zero $j^{\mu}_{\text{geom.}}$ according to our previous discussion.

\textit{Discussion:} A natural question to ask is how one can distinguish the geometric current of Eq.~\eqref{eq:1D-current}
from a more typical longitudinal current of the Drude form. The Drude contribution has the form
\beq
	j^1_{\text{Drude}}(r^1)= \tau\frac{Q^2}{h}E_1(r^1)\sum_{I=1}^{n_F}\left[v(k_{I,+})-v(k_{I,-}) \right]\ ,
\eeq
where $\tau$ is the relaxation time and $v(k)=\frac{\pd \mathcal{E}(k)}{\pd k}$. To distinguish this from Eq.~\eqref{eq:1D-current} 
we choose an electric field which is a pure gradient around an origin,
$E_1(r^1)= E^{(0)}_{11}r^1$. 
We then compute the average of the current over a 
spatial region centered at that origin $r^1\in[-\frac{L}{2},\frac{L}{2}]$. 
We find that $\int_{-\frac{L}{2}}^{\frac{L}{2}} dr^1\ j^1_{\text{Drude}}(r^1)=0$, while 
\beq
	\int_{-\frac{L}{2}}^{\frac{L}{2}}dr^1 \ j^1_{\text{geom.}}(r^1)= -\frac{L Q^2 E^{(0)}_{11}}{2h}\sum_{I=1}^{n_F}\left[ g^{11}(K_{I,+})- g^{11}(K_{I,-})\right]\ . \label{eq:avg-geom-current}
\eeq
Thus, a spatial average of the current about the origin can distinguish between these two kinds of responses when the electric
field is a pure gradient (``pure" refers to the fact that $E_1(0)=0$ and $E_1(r^1)$ is linear near $r^1=0$). 

Eq.~\eqref{eq:avg-geom-current} shows that information about the quantum metric at the Fermi points can be extracted
from a transport experiment using an electric field which is a pure gradient. By averaging the current over a spatial region
which is symmetric about the origin, any Drude contribution to the current will be canceled. Then, since
$L$ (the length of the spatial region), $Q$, $E_{11}^{(0)}$, and $h$ are known to the experimenter, the
signed sum $\sum_{I=1}^{n_F}\left[ g^{11}(K_{I,+})- g^{11}(K_{I,-})\right]$ can be extracted from this transport data.

A second natural question concerns the conditions under which the electric field gradient term is expected to 
significantly alter the semi-classical dynamics. After all, if the electric field varies slowly over the width of the wave packet, then
it should be reasonable to neglect the gradient term. To understand the relevant scales we use Eq.~\eqref{eq:quadrupole}, 
which implies that the squared spread $\lan \Psi(t)|\hat{x}^{\mu}\hat{x}^{\nu}|\Psi(t)\ran - X^{\mu}X^{\nu}$ of a wave
packet sharply peaked at $\mb{K}$ in reciprocal space is equal to $g^{\mu\nu}(\mb{K})$. 
For simplicity, consider the case of $D=1$. Then the width of the wave packet is
$\sqrt{g^{11}(\mb{K})}$ and so the change of the electric field over the width of the wave packet is 
$\Delta E_1\approx E^{(0)}_{11}\sqrt{g^{11}(\mb{K})}$. If $\Delta E_1 \ll E^{(0)}_1$ 
(the uniform part of the electric field), then we can
neglect the gradient term. On the other hand, we must include this gradient term if $\Delta E_1$ is comparable to or
larger than $E^{(0)}_1$.

\textit{Conclusion:} In this article we extended the semi-classical theory of electron wave packet
motion in solids to incorporate the effects of a non-uniform electric field. In particular, we 
systematically calculated corrections to the semi-classical EOMs in an expansion in derivatives of the electric field, and 
we obtained the correction proportional to the first derivative of the electric field.
Our main result, shown in Eqs.~\eqref{eq:new-semiclassics}, is a correction to the semi-classical EOM for the wave packet 
center in real space which depends on the electric field \emph{gradient}, and on the quantum metric 
$g^{\mu\nu}(\mb{q})$ on the BZ. 
We then gave a physical interpretation of this new term as arising from the energy associated with the electric quadrupole 
moment of the wave packet in the presence of the non-uniform electric field. 
We also showed that this correction to the semi-classical EOMs does not affect transport in insulators, 
but does lead to a nontrivial transport signature in metals with 
broken time-reversal and inversion symmetry. Specifically, we showed that in such metals an electric field gradient
can generate a longitudinal current which is proportional to the electric field gradient and to the quantum 
metric at the Fermi surface. Since the current depends only on the quantum metric at the Fermi surface, 
we expect that it will be robust to the inclusion of interaction or disorder effects, as in the case of the
anomalous Hall effect in metals~\cite{haldane2004}.

We envision at least two possible directions for future work. The first would be to understand the corrections to the
semi-classical EOMs \eqref{eq:new-semiclassics} which involve higher derivatives of the electric field. 
The correction proportional to the second derivative
would be particularly interesting as it should allow for a derivation of an analog of the formula of Hoyos and Son~\cite{hoyos-son}, 
which relates the finite wave vector Hall conductivity
of a quantum Hall system to the Hall viscosity, but in the context of Chern insulators (where there is no magnetic
field) instead of Landau levels. 
A second direction would be to derive semi-classical EOMs for the higher moments of the
wave packet, for example the second moments $X^{\mu\nu}(t):=\lan\Psi(t)|\hat{x}^{\mu}\hat{x}^{\nu}|\Psi(t)\ran$ 
and $K_{\mu\nu}(t):=\lan\Psi(t)|\hat{q}_{\mu}\hat{q}_{\nu}|\Psi(t)\ran$ in position and reciprocal space, respectively.
In particular, it would be interesting to understand how these second moments respond to non-uniform electric fields. 
We leave these topics for future work.

\textit{Note added:} After this work was completed we became aware of Ref.~\onlinecite{gao-xiao}, which obtained many
of the same results as part of a study of \emph{nonreciprocal directional dichroism} in crystalline solids.

\textit{Acknowledgments:} We thank A. Alexandradinata for a useful conversation. M.F.L. acknowledges the support of the Kadanoff 
Center for Theoretical Physics at 
the University of Chicago. T.L.H thanks the U.S. National Science Foundation under grant DMR 1351895-CAR for support.


%

\clearpage


\begin{widetext}
\section{Supplemental material for ``Semi-classical wave packet dynamics in non-uniform electric fields''}
\end{widetext}

\subsection{Quadrupole moment of the wave packet state}

Here we present an approximate calculation of the quadrupole moment 
$\lan \Psi(t)|\hat{x}^{\mu}\hat{x}^{\nu}|\Psi(t)\ran$
of the wave packet state $|\Psi(t)\ran$ introduced in the main text. Our calculation uses the assumption that the wave
packet is sharply peaked in reciprocal space. The quadrupole moment enters the calculation of 
the wave packet energy $\lan\Psi(t)|\hat{H}|\Psi(t)\ran$, since we have
\begin{align}
	\lan\Psi(t)|\hat{H}|\Psi(t)\ran= \lan\Psi(t)|\hat{H}_0|\Psi(t)\ran+ Q\lan\Psi(t)|\vphi(\hat{\mb{x}})|\Psi(t)\ran\ ,
\end{align}
and for our choice of potential the second term here takes the form
\beqa
	\lan\Psi(t)|\vphi(\hat{\mb{x}})|\Psi(t)\ran&=& -E^{(0)}_{\mu}\lan \Psi(t)|\hat{x}^{\mu}|\Psi(t)\ran \nnb \\
	 &-& \frac{1}{2}E^{(0)}_{\mu\nu}\lan \Psi(t)|\hat{x}^{\mu}\hat{x}^{\nu}|\Psi(t)\ran \ .
\eeqa
By definition of the wave packet center, we have $\lan \Psi(t)|\hat{x}^{\mu}|\Psi(t)\ran= X^{\mu}$ in the
first term here. The second term involves the quadrupole moment of the wave packet. We now turn to the calculation of the
quadrupole moment. 

Our calculation of the quadrupole moment uses two important formulas for the matrix elements of 
$\hat{x}^{\mu}$ and $\hat{x}^{\mu}\hat{x}^{\nu}$ in the Bloch states $|\psi_{\mb{q}}\ran$ which diagonalize 
$\hat{H}_0$. These formulas are:
\beq
	\lan\psi_{\mb{q}}|\hat{x}^{\mu}|\psi_{\mb{q}'}\ran= i\frac{\pd}{\pd q_{\mu}}\delta^{(D)}(\mb{q}-\mb{q}') + \delta^{(D)}(\mb{q}-\mb{q}')\mathcal{A}^{\mu}(\mb{q}) \label{eq:important-formula-1}
\eeq
and
\begin{widetext}
\begin{align}
	\lan\psi_{\mb{q}}&|\hat{x}^{\mu}\hat{x}^{\nu}|\psi_{\mb{q}'}\ran = \frac{\pd^2}{\pd q_{\mu}\pd q_{\nu}'}\delta^{(D)}(\mb{q}-\mb{q}') - i \mathcal{A}^{\nu}(\mb{q})\frac{\pd}{\pd q_{\mu}'}\delta^{(D)}(\mb{q}-\mb{q}') \nnb \\
&-  i \mathcal{A}^{\mu}(\mb{q})\frac{\pd}{\pd q_{\nu}'}\delta^{(D)}(\mb{q}-\mb{q}') -\delta^{(D)}(\mb{q}-\mb{q}')\Big\lan u_{\mb{q}}\Big| \frac{\pd^2 u_{\mb{q}}}{\pd q_{\mu}\pd q_{\nu}}\Big\ran\ .  \label{eq:important-formula-2}
\end{align}

To proceed, we first use Eq.~\eqref{eq:important-formula-2} to find the exact expression for the quadrupole moment,
\beqa
	\lan \Psi(t)|\hat{x}^{\mu}\hat{x}^{\nu}|\Psi(t)\ran= \int d^D\mb{q}\ \Bigg\{ \frac{\pd \ov{a}(\mb{q},t)}{\pd q_{\mu}}\frac{\pd a(\mb{q},t)}{\pd q_{\nu}} + i\ov{a}(\mb{q},t)\frac{\pd a(\mb{q},t)}{\pd q_{\mu}}\mathcal{A}^{\nu}(\mb{q}) \nnb \\
	+ i\ov{a}(\mb{q},t)\frac{\pd a(\mb{q},t)}{\pd q_{\nu}}\mathcal{A}^{\mu}(\mb{q})- |a(\mb{q},t)|^2 \Big\lan u_{\mb{q}}\Big| \frac{\pd^2 u_{\mb{q}}}{\pd q_{\mu}\pd q_{\nu}}\Big\ran \Bigg\}\ . \label{eq:quadrupole}
\eeqa
\end{widetext}

We are interested in the approximate evaluation of this expression in the case that the wave packet is sharply peaked 
in reciprocal space. To do this, we follow the method of Sundaram and Niu~\cite{sundaram-niu}
and first simplify the exact expression for $X^{\mu}$ under the 
assumption of a sharply peaked wave packet. Using \eqref{eq:important-formula-1} we find the exact formula
\beq
	X^{\mu}= \int d^D\mb{q}\ \left\{ i \ov{a}(\mb{q},t)\frac{\pd a(\mb{q},t)}{\pd q_{\mu}} + |a(\mb{q},t)|^2 \mathcal{A}^{\mu}(\mb{q})\right\}\ .
\eeq
Next, we multiply and divide the first term by $a(\mb{q},t)$ and use the fact that 
$|a(\mb{q},t)|^2\approx \delta^{(D)}(\mb{q}-\mb{K})$ (because the wave packet is sharply peaked in reciprocal space) 
to find the approximate expression
\beq
	X^{\mu} \approx \left[i \frac{1}{a(\mb{q},t)}\frac{\pd a(\mb{q},t)}{\pd q_{\mu}}\right]_{\mb{q}=\mb{K}} + \mathcal{A}^{\mu}(\mb{K})\ . \label{eq:approx-dipole}
\eeq
We now use this to simplify the formula for $\lan \Psi(t)|\hat{x}^{\mu}\hat{x}^{\nu}|\Psi(t)\ran$. 

We start by integrating by parts in the second term in Eq.~\eqref{eq:quadrupole} to obtain
\begin{widetext}
\beqa
	\lan \Psi(t)|\hat{x}^{\mu}\hat{x}^{\nu}|\Psi(t)\ran= \int d^D\mb{q}\ \Bigg\{ \frac{\pd \ov{a}(\mb{q},t)}{\pd q_{\mu}}\frac{\pd a(\mb{q},t)}{\pd q_{\nu}} - i\frac{\pd \ov{a}(\mb{q},t)}{\pd q_{\mu}} a(\mb{q},t)\mathcal{A}^{\nu}(\mb{q}) - i |a(\mb{q},t)|^2\frac{\pd \mathcal{A}^{\nu}(\mb{q})}{\pd q_{\mu}} \nnb \\
	+ i\ov{a}(\mb{q},t)\frac{\pd a(\mb{q},t)}{\pd q_{\nu}}\mathcal{A}^{\mu}(\mb{q})- |a(\mb{q},t)|^2 \Big\lan u_{\mb{q}}\Big| \frac{\pd^2 u_{\mb{q}}}{\pd q_{\mu}\pd q_{\nu}}\Big\ran \Bigg\}\ . 
\eeqa
This can be rewritten exactly as
\beqa
	\lan \Psi(t)|\hat{x}^{\mu}\hat{x}^{\nu}|\Psi(t)\ran &=& \int d^D\mb{q}\ \Bigg\{ \left(\frac{\pd \ov{a}(\mb{q},t)}{\pd q_{\mu}} + i\mathcal{A}^{\mu}(\mb{q})\ov{a}(\mb{q},t)\right)\left(\frac{\pd a(\mb{q},t)}{\pd q_{\nu}} - i\mathcal{A}^{\nu}(\mb{q})a(\mb{q},t)\right) - |a(\mb{q},t)|^2 \mathcal{A}^{\mu}(\mb{q})\mathcal{A}^{\nu}(\mb{q}) \nnb \\
	&-&  i |a(\mb{q},t)|^2\frac{\pd \mathcal{A}^{\nu}(\mb{q})}{\pd q_{\mu}} - |a(\mb{q},t)|^2 \Big\lan u_{\mb{q}}\Big| \frac{\pd^2 u_{\mb{q}}}{\pd q_{\mu}\pd q_{\nu}}\Big\ran \Bigg\}\ .
\eeqa
Next, we can write
\beqa
	\Big\lan u_{\mb{q}}\Big| \frac{\pd^2 u_{\mb{q}}}{\pd q_{\mu}\pd q_{\nu}}\Big\ran &=& \frac{\pd}{\pd q_{\mu}}\left( \Big\lan u_{\mb{q}}\Big| \frac{\pd u_{\mb{q}}}{\pd q_{\nu}}\Big\ran\right) - \Big\lan \frac{\pd u_{\mb{q}}}{\pd q_{\mu}}\Big| \frac{\pd u_{\mb{q}}}{\pd q_{\nu}}\Big\ran \nnb \\
	&=& -i \frac{\pd \mathcal{A}^{\nu}(\mb{q})}{\pd q_{\mu}} - \Big\lan \frac{\pd u_{\mb{q}}}{\pd q_{\mu}}\Big| \frac{\pd u_{\mb{q}}}{\pd q_{\nu}}\Big\ran \ .
\eeqa
Plugging this back into our expression for the quadrupole moment gives
\beqa
	\lan \Psi(t)|\hat{x}^{\mu}\hat{x}^{\nu}|\Psi(t)\ran &=& \int d^D\mb{q}\ \Bigg\{ \left(\frac{\pd \ov{a}(\mb{q},t)}{\pd q_{\mu}} + i\mathcal{A}^{\mu}(\mb{q})\ov{a}(\mb{q},t)\right)\left(\frac{\pd a(\mb{q},t)}{\pd q_{\nu}} - i\mathcal{A}^{\nu}(\mb{q})a(\mb{q},t)\right) - |a(\mb{q},t)|^2 \mathcal{A}^{\mu}(\mb{q})\mathcal{A}^{\nu}(\mb{q}) \nnb \\
	&+& |a(\mb{q},t)|^2 \Big\lan \frac{\pd u_{\mb{q}}}{\pd q_{\mu}}\Big| \frac{\pd u_{\mb{q}}}{\pd q_{\nu}}\Big\ran \Bigg\}\ .
\eeqa
We now multiply and divide the first term here by $|a(\mb{q},t)|^2$, use 
$|a(\mb{q},t)|^2\approx \delta^{(D)}(\mb{q}-\mb{K})$, and use Eq.~\eqref{eq:approx-dipole} and its complex conjugate to find that
\beq
	\lan \Psi(t)|\hat{x}^{\mu}\hat{x}^{\nu}|\Psi(t)\ran\approx X^{\mu}X^{\nu}- \mathcal{A}^{\mu}(\mb{K})\mathcal{A}^{\nu}(\mb{K}) + \Big\lan \frac{\pd u_{\mb{K}}}{\pd K_{\mu}}\Big| \frac{\pd u_{\mb{K}}}{\pd K_{\nu}}\Big\ran\ .
\eeq
Now we can write
\beqa
	 \mathcal{A}^{\mu}(\mb{K})\mathcal{A}^{\nu}(\mb{K}) &=&  - \Big\lan u_{\mb{K}}\Big| \frac{\pd u_{\mb{K}}}{\pd K_{\mu}}\Big\ran \Big\lan u_{\mb{K}}\Big| \frac{\pd u_{\mb{K}}}{\pd K_{\nu}}\Big\ran \nnb \\
	 &=& \Big\lan \frac{\pd u_{\mb{K}}}{\pd K_{\mu}} \Big| u_{\mb{K}}\Big\ran \Big\lan u_{\mb{K}}\Big| \frac{\pd u_{\mb{K}}}{\pd K_{\nu}}\Big\ran\ ,
\eeqa
and so we find that
\beq
	\lan \Psi(t)|\hat{x}^{\mu}\hat{x}^{\nu}|\Psi(t)\ran\approx X^{\mu}X^{\nu} + \Big\lan \frac{\pd u_{\mb{K}}}{\pd K_{\mu}}\Big| \frac{\pd u_{\mb{K}}}{\pd K_{\nu}}\Big\ran - \Big\lan \frac{\pd u_{\mb{K}}}{\pd K_{\mu}} \Big| u_{\mb{K}}\Big\ran \Big\lan u_{\mb{K}}\Big| \frac{\pd u_{\mb{K}}}{\pd K_{\nu}}\Big\ran\ .
\eeq
\end{widetext}
We have nearly arrived at the final answer. The last ingredient is to return to the original expression 
Eq.~\eqref{eq:quadrupole} and note that since
\beq
	\int d^D\mb{q}\ \frac{\pd \ov{a}(\mb{q},t)}{\pd q_{\mu}}\frac{\pd a(\mb{q},t)}{\pd q_{\nu}}= \int d^D\mb{q}\ \frac{\pd \ov{a}(\mb{q},t)}{\pd q_{\nu}}\frac{\pd a(\mb{q},t)}{\pd q_{\mu}}\ ,
\eeq
(to see it, integrate by parts twice to exchange the derivatives) we could have replaced the original integral expression for 
$\lan \Psi(t)|\hat{x}^{\mu}\hat{x}^{\nu}|\Psi(t)\ran$ with a 
symmetrized version of it, 
$\lan \Psi(t)|\hat{x}^{\mu}\hat{x}^{\nu}|\Psi(t)\ran= \frac{1}{2}[$Eq.~\eqref{eq:quadrupole}$+(\mu \leftrightarrow \nu)]$. Going through analogous manipulations for this symmetrized expression then yields our final
expression for the quadrupole moment of $|\Psi(t)\ran$,
\beq
	\lan \Psi(t)|\hat{x}^{\mu}\hat{x}^{\nu}|\Psi(t)\ran\approx X^{\mu}X^{\nu} + g^{\mu\nu}(\mb{K})\ .
\eeq

We see that this expression involves the quantum metric evaluated at the wave packet center in reciprocal space. 
In addition, we find that the connected part of the quadrupole moment is given by
\beq
	\lan \Psi(t)|\hat{x}^{\mu}\hat{x}^{\nu}|\Psi(t)\ran - \lan \Psi(t)|\hat{x}^{\mu}|\Psi(t)\ran \lan \Psi(t)|\hat{x}^{\nu}|\Psi(t)\ran \approx  g^{\mu\nu}(\mb{K})\ \label{eq:connected-x2}.
\eeq
We learn from this expression that if we assume that the wave packet is sharply peaked at position $\mb{K}$
in reciprocal space, then the spread of the wave packet in real space --- which is measured by the connected part of the
quadrupole moment --- is given by the quantum metric at the location $\mb{K}$.

If we use our result for the quadrupole moment of the wave packet, then we find that the total wave packet energy is given
by the approximate expression  
\begin{align}
\lan\Psi(t)|\hat{H}|\Psi(t)\ran &\approx \mathcal{E}(\mb{K}) - Q E^{(0)}_{\mu}X^{\mu} \nnb \\
&- \frac{1}{2}Q E^{(0)}_{\mu\nu}\left( X^{\mu}X^{\nu} + g^{\mu\nu}(\mb{K}) \right)\ .
\end{align}
This is the quantity that we called $\mathcal{E}_{eff}(\mb{X},\mb{K})$ in the main text.

\subsection{Issues related to possible boundary terms}

In the main text we noted that in our derivation of Eqs.~3 and 5 (of the main text) we integrated by parts in integrals over the
Brillouin zone (BZ) and we neglected boundary terms. In this section we prove that in a suitable choice of gauge for the Bloch states 
$|\psi_{\mb{q}}\ran$, namely the \emph{periodic gauge}, the wave packet amplitude $a(\mb{q},t)$ can be chosen to be a periodic function 
on the BZ. This means that in this gauge the only possible boundary terms which could appear in the derivation of Eqs.~3 and 5 are
terms associated with the Berry connection $\mathcal{A}^{\mu}(\mb{q})$. It is well-known that for certain band
structures the Berry connection for a given band may not be periodic on the BZ, and this might lead to nontrivial boundary terms which
would appear as corrections to Eqs.~3 and 5 of the main text. We leave an exploration of such boundary terms for future work. 

The \emph{periodic gauge} for the Bloch states $|\psi_{\mb{q}}\ran$ is defined by the condition
\beq
	|\psi_{\mb{q}+\mb{G}}\ran=|\psi_{\mb{q}}\ran \label{eq:periodic-gauge}
\eeq
for all reciprocal lattice vectors $\mb{G}$ (see, for example, Ref.~\onlinecite{king-smith}). Recall also that the
wave packet state that we consider in the main text has the form
\beq
	|\Psi(t)\ran= \int d^D\mb{q}\ a(\mb{q},t)|\psi_{\mb{q}}\ran\ .
\eeq
We now prove the following. Suppose that we adopt the periodic gauge for the Bloch states $|\psi_{\mb{q}}\ran$, and
we also choose the wave packet amplitude $a(\mb{q},0)$ at time $t=0$ to obey the periodicity condition
\beq
	a(\mb{q}+\mb{G},0)=a(\mb{q},0) 
\eeq
for all reciprocal lattice vectors $\mb{G}$. Then at all later times $t>0$ the amplitude $a(\mb{q},t)$ remains periodic,
i.e., we have
\beq
	a(\mb{q}+\mb{G},t)=a(\mb{q},t) 
\eeq
for all reciprocal lattice vectors $\mb{G}$ and for all $t>0$. 

To prove this we first recall the equation of motion for $a(\mb{q},t)$,
\begin{align}
	i\hbar\dot{a}(\mb{q},t)= a(\mb{q},t)\mathcal{E}(\mb{q}) + Q \int d^D\mb{q}'\ a(\mb{q}',t) \lan\psi_{\mb{q}}|\vphi(\hat{\mb{x}})|\psi_{\mb{q}'}\ran\ . 
\end{align}
The equation of motion for $a(\mb{q}+\mb{G},t)$ is then
\begin{align}
	i\hbar\dot{a}(\mb{q}+\mb{G},t)&= a(\mb{q}+\mb{G},t)\mathcal{E}(\mb{q}+\mb{G}) \nnb \\
+&\ Q \int d^D\mb{q}'\ a(\mb{q}',t) \lan\psi_{\mb{q}+\mb{G}}|\vphi(\hat{\mb{x}})|\psi_{\mb{q}'}\ran\ . 
\end{align}
Since $\mb{G}$ is a reciprocal lattice vector we always have
\beq
	\mathcal{E}(\mb{q}+\mb{G})=\mathcal{E}(\mb{q})
\eeq
since $\mathcal{E}(\mb{q})$, the energy of the Bloch state $|\psi_{\mb{q}}\ran$, is periodic in reciprocal space. In
addition, in the periodic gauge we also have Eq.~\eqref{eq:periodic-gauge}. This means that in the 
periodic gauge the equation of motion for $a(\mb{q}+\mb{G},t)$ takes the form
\begin{align}
	i\hbar\dot{a}(\mb{q}+\mb{G},t)&= a(\mb{q}+\mb{G},t)\mathcal{E}(\mb{q}) \nnb \\
+&\ Q \int d^D\mb{q}'\ a(\mb{q}',t) \lan\psi_{\mb{q}}|\vphi(\hat{\mb{x}})|\psi_{\mb{q}'}\ran\ . 
\end{align}
Now let us define the quantity
\beq
	b(\mb{q},t):= a(\mb{q}+\mb{G},t)-a(\mb{q},t)\ .
\eeq
By subtracting the equations of motion for $a(\mb{q}+\mb{G},t)$ and $a(\mb{q},t)$, we find that
$b(\mb{q},t)$ evolves in time according to the simple equation
\beq
	i\hbar\dot{b}(\mb{q},t)= b(\mb{q},t)\mathcal{E}(\mb{q})
\eeq
with solution
\beq
	b(\mb{q},t)= e^{-i\frac{\mathcal{E}(\mb{q})t}{\hbar}}b(\mb{q},0)\ .
\eeq
It follows that if we choose $b(\mb{q},0)= a(\mb{q}+\mb{G},0)-a(\mb{q},0)=0$, then we have
$b(\mb{q},t)= a(\mb{q}+\mb{G},t)-a(\mb{q},t)=0$ for all later times $t$, and this completes the proof.

\subsection{Details of the example lattice models displaying a nontrivial geometric current}

\begin{figure}[t!]
  \centering
    \includegraphics[width= .5\textwidth]{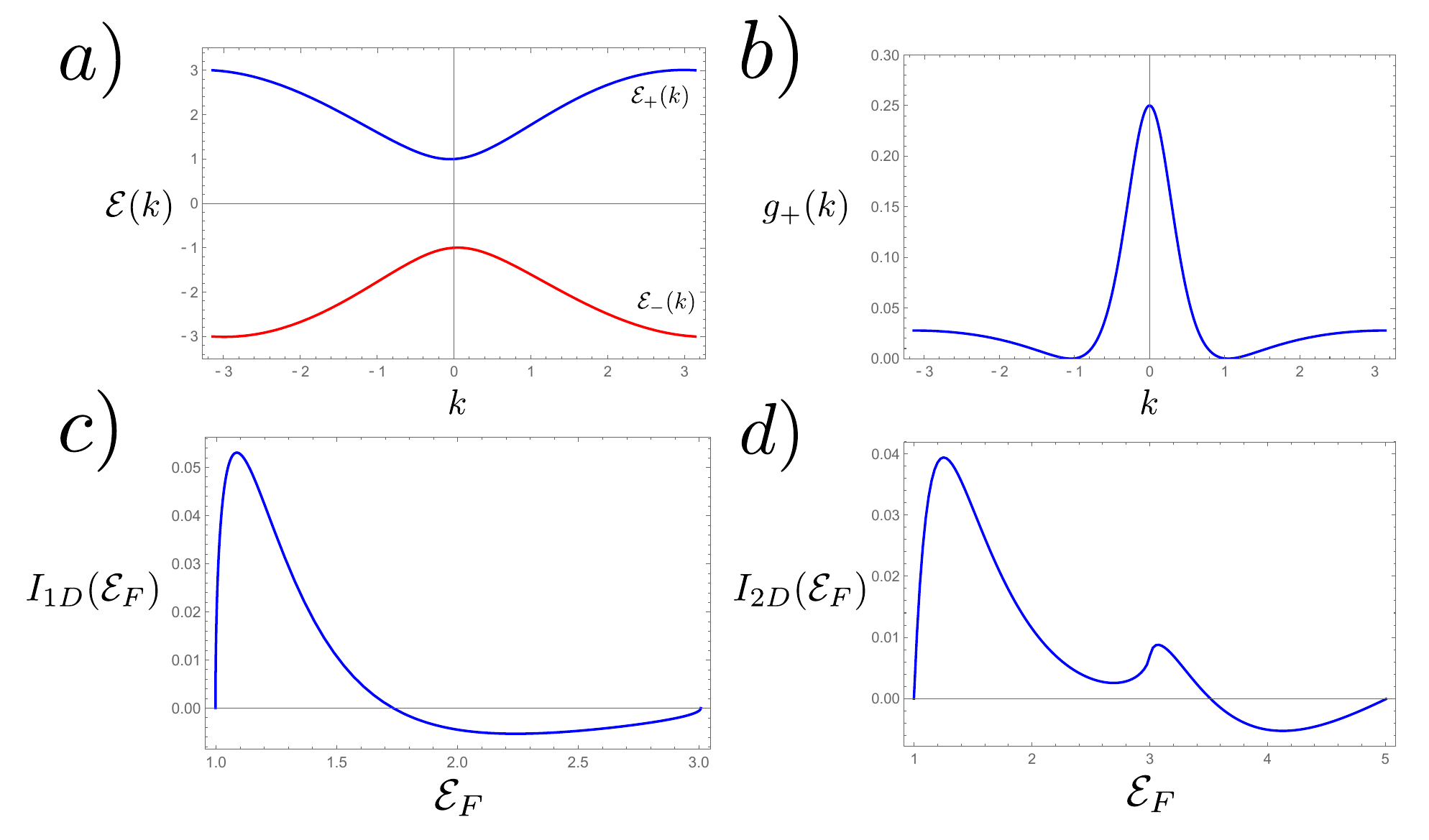} 
 \caption{ a) The band energies $\mathcal{E}_{\pm}(k)$ for the two bands of the 1D model 
 Eq.~\eqref{eq:example-ham}. 
 b) The quantum metric $g_{+}(k)$ for the upper band of the 1D model Eq.~\eqref{eq:example-ham}. c) The function 
$I_{1D}(\mathcal{E}_F)$ from Eq.~\eqref{eq:1D-integral} for the 1D model Eq.~\eqref{eq:example-ham}. d) The function 
$I_{2D}(\mathcal{E}_F)$ from Eq.~\eqref{eq:2D-integral} for the 2D model Eq.~\eqref{eq:example-ham-2D}. 
For all of the plots the parameter values are chosen to be $m=1$, $A=0.1$. In plots c) and d) the values of $\mathcal{E}_F$ on the 
horizontal axis range all the way from the bottom to the top of the conduction band (the upper band in both models).}
\label{fig1}
\end{figure}

In this section we present the details of the calculation of the geometric current in the two example lattice models that we mentioned in the
main text. 

\subsubsection{An example in $D=1$}

We start with an example of a lattice model in $D=1$ which yields a non-zero $j^1_{\text{geom.}}$. 
We consider the two-band model with Bloch Hamiltonian (we write $k\equiv k_1$ since we are in $D=1$)
\beq
	H(k)= A\sin(k)\mathbb{I} + \sin(k)\sigma^x + (m+1-\cos(k))\sigma^z\ , \label{eq:example-ham}
\eeq
where $\mathbb{I}$ is the $2\times 2$ identity matrix and $\sigma^{x,y,z}$ are the Pauli matrices. We assume that
the parameters $m$ and $A$ are chosen so that there is an energy gap between the two bands of the model. To obtain
a model of a metal we then completely fill the lower band and partially fill the upper band to a Fermi energy
$\mathcal{E}_F$ (we work at zero temperature). For $A=0$ this model
has inversion and time-reversal symmetry, which are both given by $H(-k)=\sigma^z H(k)\sigma^z$ since $H(k)$ is real. Thus,
according to the symmetry analysis in the main text, we need $A\neq 0$ to obtain a non-zero geometric current. The
band energies for this model are given by $\mathcal{E}_{\pm}(k)= \pm \lambda(k) + A\sin(k)$ with
$\lambda(k)= \sqrt{\sin^2(k)+(m+1-\cos(k))^2}$, and the quantum metric for the upper band is given by
\beq
	g_{+}(k)= \frac{\left[(1+m)\cos(k)-1\right]^2}{4\lambda(k)^4}\ .
\eeq
In particular, $g_{+}(k)$ is independent of $A$ as the identity matrix term in $H(k)$ does not change the
form of the eigenvectors of $H(k)$. 

We now show that this model yields a non-zero $j^1_{\text{geom.}}$. 
Let $I_{1D}(\mathcal{E}_F)$ denote the integral
\beq
	I_{1D}(\mathcal{E}_F)= \int dk\ f_0(k) \frac{\pd g_{+}(k)}{\pd k}\ . \label{eq:1D-integral}
\eeq
This integral is a function of the Fermi energy $\mathcal{E}_F$, since $f_0(k)$, the equilibrium distribution function at zero 
temperature, is completely determined by $\mathcal{E}_F$. From Eq.~11 of the main text, the geometric current in the wire in 
the presence of an electric field gradient will be proportional to $I_{1D}(\mathcal{E}_F)$. For the parameter
values $m=1$, $A=0.1$, we plot the function $I_{1D}(\mathcal{E}_F)$ in Fig.~\ref{fig1}c for a range of values of 
$\mathcal{E}_F$ which span the entire ``$+$'' band of the model. We find that $I_{1D}(\mathcal{E}_F)\neq 0$ in general, and 
so this model has $j^1_{\text{geom.}}\neq 0$ for generic values of $\mathcal{E}_F$. For reference, the band 
energies $\mathcal{E}_{\pm}(k)$ and the quantum metric $g_{+}(k)$ for the parameter values $m=1$, $A=0.1$ are plotted
in Fig.~\ref{fig1}a and Fig.~\ref{fig1}b.

\subsubsection{An example in $D=2$} 

We now briefly discuss a model in $D=2$ which yields a non-zero 
$j^{\mu}_{\text{geom.}}$. This model has a Bloch Hamiltonian given by
\begin{align}
	H(\mb{k})&= A\sin(k_1)\mathbb{I} + \sin(k_1)\sigma^x + \sin(k_2)\sigma^y \nnb \\
	&+ (m+2-\cos(k_1)-\cos(k_2))\sigma^z \ . \label{eq:example-ham-2D}
\end{align}
For $A=0$ this model can describe a Chern insulator on the square lattice when the lower band is full and 
$-4<m<-2$ or $-2<m<0$. Here we consider the model for $m>0$, and we fill the lower band completely and only partially
fill the upper band to obtain a model of a metal. For $m>0$ the upper band has a minimum at $\mb{k}=(0,0)$ and
maxima at the BZ corners $\mb{k}=(\pm\pi,\pm\pi)$. The quantum metric for this model is independent of $A$
and is given in Eq.~62 of Ref.~\onlinecite{matsuura2010momentum}. For $A=0$ this model has inversion symmetry
given by $H(-\mb{k})=\sigma^z H(\mb{k})\sigma^z$, and so we need $A\neq 0$ to have $j^{\mu}_{\text{geom.}}\neq 0$. To illustrate 
the nontrivial response we consider a non-uniform electric
field in the $X^1$ direction only, so we only turn on $E^{(0)}_{11}\neq 0$. As a result, we only need to study the
term in $j^1_{\text{geom.}}$ which contains $g^{11}(\mb{k})$. In this case, 
$j^1_{\text{geom.}}$ will be proportional to the integral
\beq
	I_{2D}(\mathcal{E}_F)= \int d^2\mb{k}\ f_0(\mb{k}) \frac{\pd g^{11}(\mb{k})}{\pd k_1}\ , 
	\label{eq:2D-integral}
\eeq
where $f_0(\mb{k})$ again denotes the zero temperature distribution function determined by
$\mathcal{E}_F$. We plot $I_{2D}(\mathcal{E}_F)$ in Fig.~\ref{fig1}d for a range of $\mathcal{E}_F$ values which stretches
from the bottom to the top of the upper band in this model. The plot again shows that $I_{2D}(\mathcal{E}_F)\neq 0$ in
general, which implies a non-zero $j^1_{\text{geom.}}$ for generic values of $\mathcal{E}_F$.

\end{document}